# Optimization of Laser Irradiation Uniformity for the Double-Cone Ignition Scheme with MULTI-3D simulations

Yicheng Wang[1,2], Yiwen Yang[1,2], Fuyuan Wu[1,3], Yuhan Wang[1,3], Rafael Ramis[4]，Jie Zhang[1,2,3]

1）School of Physics and Astronomy, Shanghai Jiao Tong University, Shanghai 200240,China
2）Zhiyuan College, Shanghai Jiao Tong University, Shanghai 200240, China
3）State Key Laboratory of Dark Matter Physics, Key Laboratory for Laser Plasmas (MOE), Shanghai Jiao Tong University, Shanghai 200240, China
4）E.T.S.I. Aeronautica y del Espacio, Universidad Politecnica de Madrid, Madrid 28040, Spain

**Abstract**: The double-cone ignition (DCI) scheme holds a promising perspective for laser driven fusion energy and astrophysics. However, optimizing the laser irradiation uniformity under the constraints of limited laser beams and a given cone angle remains to be explored. We utilized the three-dimensional radiation hydrodynamics program MULTI-3D to simulate the interaction process between the laser and plasma shell. By employing Bayesian optimization for the pointing position of the incident laser beams, we achieved a laser irradiation scheme with nonuniformity less than 5%. This study can provide references for experiments and offer valuable insights for other laser fusion schemes.
**Keywords:** Laser fusion; Irradiation uniformity; Hydrodynamic simulation; Bayesian optimization.

‡ Yicheng Wang and Yiwen Yang contributed equally to this work.
Authors to whom correspondence should be addressed: jzhang1@sjtu.edu.cn and fuyuan.wu@sjtu.edu.cn

## 1 Introduction

Laser fusion is a strong candidate for achieving low-carbon energy, playing a significant role in the carbon neutrality efforts to protect the Earth's environment [1-5]. In recent years, important progress has been made in laser fusion research. In 2022, the National Ignition Facility (NIF) in the United States achieved a fusion ignition target with an energy gain greater than 1 for the first time, heralded as the "Wright Brothers moment" for fusion energy [2]. In 2024, the University of Rochester achieved a thermal ignition target with a fusion fuel gain greater than unit [5]. However, laser fusion schemes aimed at fusion power generation require even higher fusion gain and laser energy utilization efficiency.

The double-cone ignition (DCI) scheme combines the advantages of several ignition schemes including laser direct drive, fast electron ignition, and magnetic field assisted fusion, promising efficient ignition and high-gain burning [3,6-9]. Additionally, the DCI scheme can high-density degenerated plasma jets, providing an excellent platform for the study of the astrophysical of compact objects [9].

Recently, the upgraded ShenGuang-II Laser Facility underwent another upgrade, changing the driving laser from 8 beams to 16 beams. Therefore, it is essential to study the laser irradiation uniformity in the new configuration. The laser irradiation uniformity on the target surface significantly impacts the ablation and compression of the plasma. The DCI scheme does not rely on central hotspot ignition and exhibits strong resistance to Rayleigh-Taylor instability. However, if the laser irradiation uniformity is too poor, it could still lead to severe fluid instability and implosion asymmetry, reducing the plasma density and fusion performance [10,11].

The laser irradiation nonuniformity is a fundamental problem that needs to be addressed for direct-drive fusion [12-16]. Over the past few decades, many prominent researchers have studied this issue. For instance, Temporal et al. examined laser irradiation uniformity in shock ignition scheme [12], while Murakami et al. focused on it in central ignition schemes [13], and Ramis et al. explored it in the polarization drive scheme of the LMJ facility [14]. However, there has been less research on laser irradiation uniformity under gold cone conditions. This paper employs a combination of machine learning and three-dimensional radiation fluid simulations for the first time to optimize the laser irradiation uniformity in the double cone ignition scheme. Compared to traditional ray geometry tracing methods, the radiation fluid dynamics approach can take the effects of incident laser and plasma motion on irradiation uniformity into account simultaneously, providing a more reliable reference for experiments.

This paper is divided into five sections. The second section introduces the simulation program and the optimization method with machine learning. The third section presents the optimization results and physical analysis. The fourth section provides synthesized images and their analysis. Finally, the conclusions and discussion of the study are presented.

**2 Physical model, simulation program and optimization method**

The MULTI-3D program is a radiation fluid dynamics code widely used in research areas such as laser fusion, Z-pinch fusion, and heavy ion fusion [17-21]. The advantages of the MULTI-3D program include its ability to handle complex geometric models, multi-density, as well as various optical thickness models. The program achieves stability in the matter-energy-radiation coupling system in two ways. One is employing the symmetric semi-implicit method. Another one is discretizing radiation frequencies and positions with an unstructured three-dimensional mesh composed of tetrahedral elements. However, MULTI-3D also has some limitations [21]. For instance, it currently solves the radiation fluid dynamics equations solely within a Lagrangian framework, restricting its ability to simulate significant grid distortion and the full implosion process in DCI schemes.

The control equations of the MULTI-3D are as following:

$$\rho = \frac{dm}{dV} \quad (1)$$

$$\rho \frac{du}{dt} = \nabla(P + P_v) \quad (2)$$

$$\rho \frac{de}{dt} = -(P + P_v)\nabla \cdot u - \nabla \cdot (q_e) + Q_{laser} + Q_{rad} \quad (3)$$

$$n \cdot \nabla I = \frac{I^P - I}{\lambda} \quad (4)$$

Equations (1)-(3) represent the conservation of mass, momentum, and energy, respectively, while Equation (4) describes the radiation transportation process. In these equations, $\rho$ is the material density, $m$ is mass, $V$ is volume, $u$ is velocity, $P$ is the thermal pressure of the material, $P_v$ is the artificial viscous pressure, $e$ is the specific internal energy, $q_e$ is the electron heat flux, $Q_{laser}$ is the laser deposition energy, $Q_{rad}$ is the radiation source, $n$ is the unit vector in direction, and $I$ is the radiation intensity, which depends on the specified unit vector and spatial position. $I^P$ denotes the blackbody radiation intensity. In systems at local thermal equilibrium, it is approximated that the radiation source in the material field follows a Planck distribution. Lastly, $\lambda$ is the mean free path of the radiation.

To optimize the laser parameters, a Bayesian optimization algorithm is implemented to call the MULTI-3D program. The Bayesian optimization algorithm sets the laser parameters as independent variables and gets the simulation results from MULTI-3D. The Bayesian posterior probability distribution of optimization possibilities in the parameter space is refreshed based on the results. New parameters are chosen according to the new

Bayesian probability. The MULTI-3D program provides the laser irradiation uniformity for the specified laser parameters.

Figure 1 illustrates the flowchart for optimizing calculations by employing the Bayesian optimization algorithm to call the MULTI-3D program. For detailed information on the implementation of the Bayesian algorithm, please refer to the pymulti4fusion library [22].

(a) Set the algorithm parameters and the ranges for the laser parameters. To accelerate the optimization process, we choose to simulate multiple sampling data and compute laser irradiation uniformity concurrently. The laser parameters include the focus position of the laser beam.

(b) Initialize the Bayesian prior probability for the optimal laser parameters and set the initial sampling laser parameters.

(c) Call the MULTI-3D program to simulate the ablation process of the target driven by the laser. Calculate the corresponding laser irradiation uniformity and other parameters.

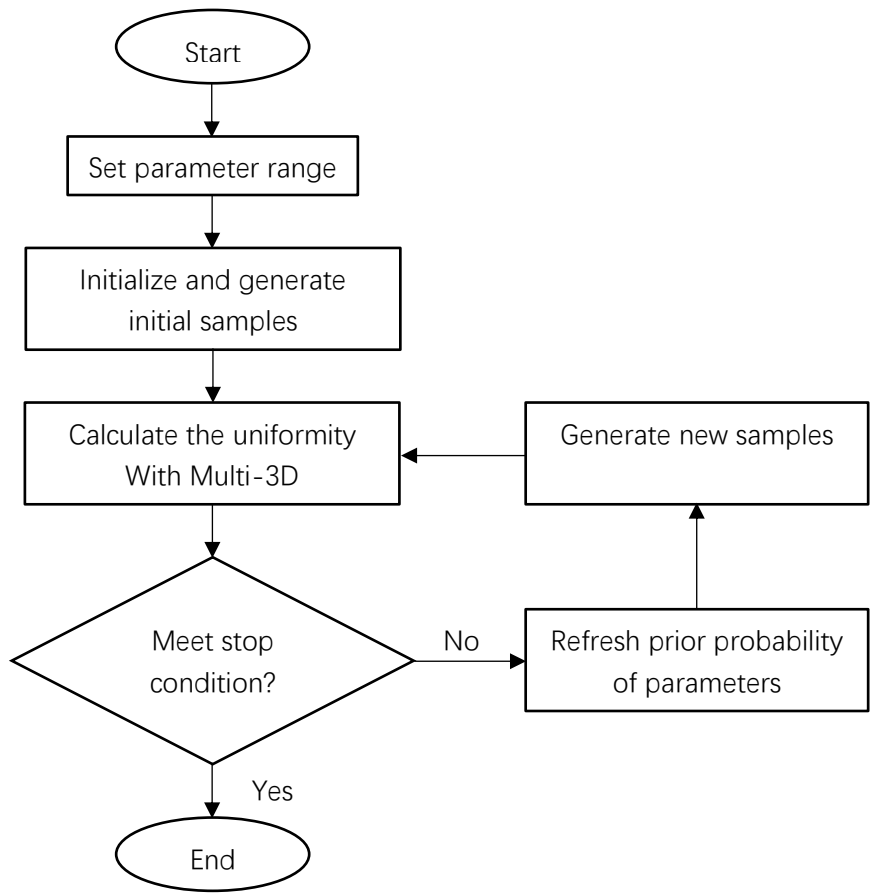

Figure 1: Optimization process of Bayesian algorithm calling MULTI-3D program

(d) Determine whether to stop the optimization based on the results calculated by the MULTI-3D program. The stopping conditions for the Bayesian optimization algorithm include reaching the maximum number of iterations or achieving convergence for the laser irradiation uniformity of the best sample.

(e) Generate new sampling laser parameters through the Bayesian optimization algorithm. The outline of the Bayesian optimization operation is as follows: the posterior probability distribution of the best laser parameters obtained from the previous round is treated as the Bayesian prior probability distribution. Based on the distribution, new points are sampled where they are most likely to be the best parameters. Those data are again used to refresh the Bayesian posterior probability distribution.

## 3 Optimize results and analysis

We used the MULTI-3D program to simulate the laser ablation and compression process of the fuel of the DCI scheme, which was carried out using the upgraded "ShenGuang-II" facility. As shown in Figure 2, the upgraded "ShenGuang-II" facility includes 16 laser beams. The opening angle of the gold cone is 100 degrees. The material of the conical target is CH. The inner radius is 450 micrometers, and the outer radius is 550 micrometers. Each conical target inside the gold cone is illuminated by double rings of laser, with four beams for each ring. The incident angle of the inner ring laser is 28 degrees, and the incident angle of the outer ring laser is 50 degrees. Different laser rings use different CPP focal spots and aiming positions. Due to the symmetry between the two gold cones, we will only analyze the laser irradiation uniformity in a single gold cone below.

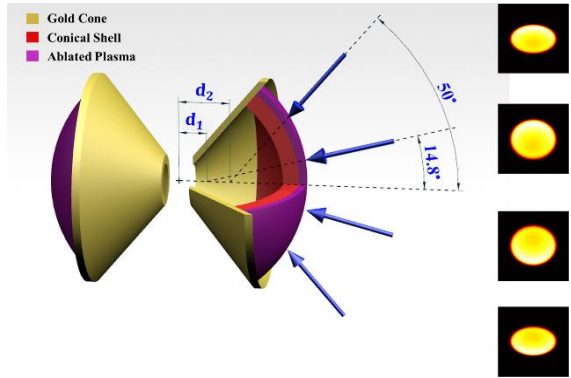

Figure 2: Schematic diagram of laser irradiation and plasma position, and the right four pictures show irradiation effect of singular laser at corresponding position.

Figure 3 shows the spatiotemporal evolution of the plasma under typical parameters. As shown in Figure 2, the aiming distances of the inner and outer ring lasers are $d_1 = 330\mu m$ and $d_2 = 346.5\mu m$, respectively. When the laser irradiates the plasma, it interacts with plasma outside the critical density surface, transferring energy to the plasma through several mechanisms such as electron-ion collisions and plasma oscillations. After gaining energy, the plasma undergoes a process of temperature rise and volume expansion. After a certain period, the ablated plasma reaches very high temperatures and pressures. On one hand, this pressure causes the coronal plasma to expand outward. On the other hand, the ablated pressure drives the high-density plasma to implode toward the spherical center. Due to the volume expansion and the partial conversion of internal energy into implosion kinetic energy, a downward trend in the highest plasma temperature can be observed.

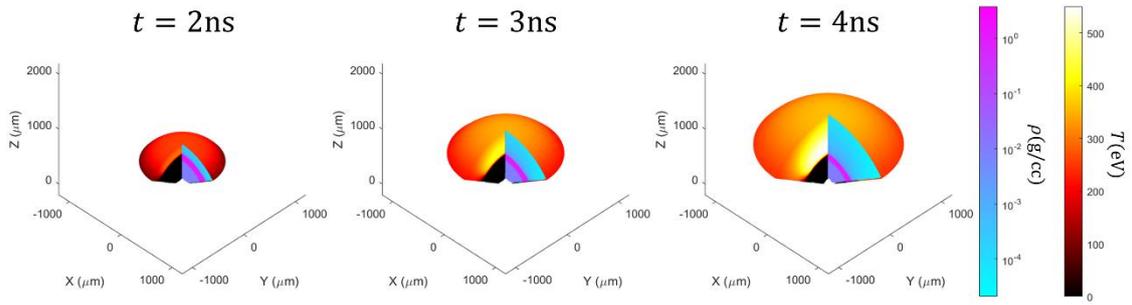

Figure 3: Plasma temperature and density distributions at different times

Figure 4 shows the distribution of plasma density, temperature, and laser energy deposition as a function of radius at two typical moments. Since the laser cannot propagate into the dense plasma regions where the density exceeds the critical density ($n_c = 1.1 \times 10^{21} \left(\frac{1}{\lambda_L}\right)^2 \left[\frac{1}{\text{cm}^3}\right]$, where $\lambda_L$ is the wavelength of laser), the laser energy deposited near the critical surface ablates the dense plasma through electron heat conduction. Therefore, the electron heat conduction region between the critical surface and the ablation surface can further smooth out the non-uniformity of the incident laser. As the critical surface changes little in the later stages of the implosion, the laser irradiation conditions remain essentially unchanged, meaning our simulation only needs to cover the early stage of the laser-plasma interaction process.

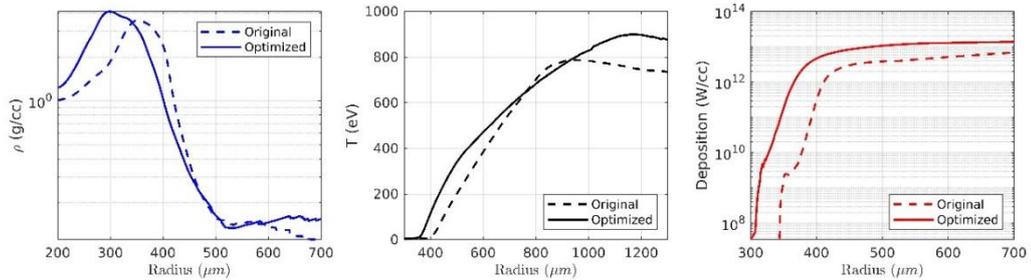

Figure 4: Distribution of plasma density, temperature, and laser energy deposition before and after optimization, parameter respectively be

$(203\mu m, 157\mu m), (267\mu m, 100\ \mu m)$

Figure 5 shows the distribution of integrated laser energy deposition $I_{depo}(\theta,\phi)$ in the longitudinal and latitudinal directions under two different parameter settings. Integrated laser energy deposition is defined in Eq. (5) as the integration of the time derivative of specific internal energy $e(r)$ caused by laser along the half-infinite line. For each half-infinite line starting from the center of the sphere (corresponding to a pair of azimuthal and polar angle $\theta,\phi$, and characterized by the unit vector $\hat{\boldsymbol{n}}(\theta,\phi)$),

$$I_{depo}(\theta,\phi) = \int_0^\infty dr \left(\frac{de}{dt}(r\hat{\boldsymbol{n}}(\theta,\phi))\right)_{laser} \tag{5}$$

From Figure 5, it can be observed that the plasma has higher energy deposition uniformity in the longitudinal direction, but the uniformity in the latitudinal direction is relatively poor. This is because there are four laser beams in the longitudinal direction, but only two laser rings in the latitudinal direction. Specifically, since the top of the spherical shell target is irradiated by four laser beams, while the edges of the target are only irradiated by one or two beams, the laser irradiation intensity at the top of the target is significantly higher than at the edges. To further optimize the irradiation uniformity in the latitudinal direction, we employed the Bayesian Optimization to optimize the aiming position of the incident laser beams.

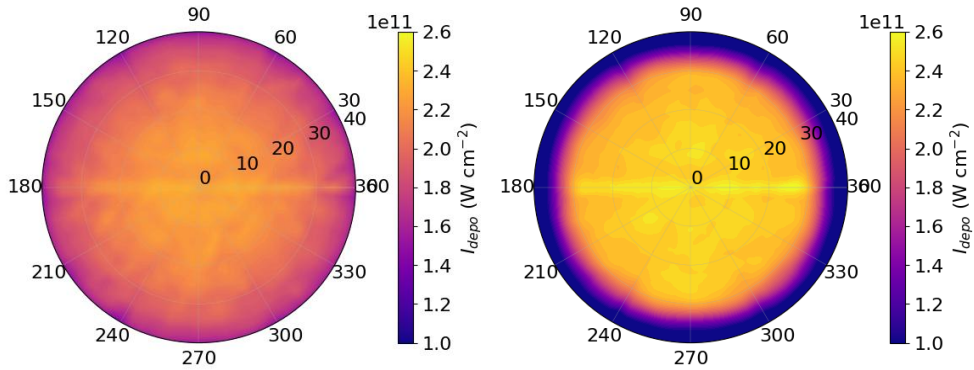

Figure 5: Distribution of integrated laser energy deposition $I_{depo}$. Left: distribution after optimization, $(d_1, d_2) = (267\mu m, 100\mu m)$. Right: typical distribution before optimization, $(d_1, d_2) = (550\mu m, 528\mu m)$.

Here, we primarily optimize the aiming distances $d_1$ and $d_2$ of the two laser rings, with the parameter range of Bayesian Optimization for both distances being $[0, 500\mu m]$. The evaluation function (which we called the uniformity $U_{depo}$) used is the variance of integrated laser energy deposition $I_{depo}$,

$$U_{depo} = \frac{1}{\bar{I}_{depo}} \left( \int_{\Omega_{sample}} d\Omega(\theta,\phi) \left[ I_{depo}(\theta,\phi) - \bar{I}_{depo} \right]^2 \right)^{1/2} \tag{6}$$

Where $\bar{I}_{depo} = \int_{\Omega_{sample}} d\Omega(\theta,\phi) I_{depo}(\theta,\phi)$, and $\Omega_{sample}$ is the sampling solid angle. To avoid sampling in gold cones, here we choose the sampling region to be $0 \leq \theta \leq 40°$. Using the Bayesian Optimization, we obtained the evaluation function variation map in the selected parameter space, which is shown in Figure 6. We can observe that the evaluation function reaches its minimum, representing the optimal parameters, at $(d_1, d_2) = (267\mu m, 100\mu m)$. As parameter $d_1$ increases and parameter $d_2$ spreads to both sides, leading to a decline in irradiation uniformity.

In the Figure 6, one point represents the aiming parameters used before Bayesian Optimization (marked by a black ×), and the other represents the optimal parameters obtained after optimization (marked by a red ×). By comparing the two, we can see that the optimal parameters, compared to the general parameters, result in a more uniform temperature distribution at the boundaries (where $\theta > 45°$) and a flatter surface at the maximum temperature. This leads to the evaluation function reaching its minimum value, achieving optimal irradiation.

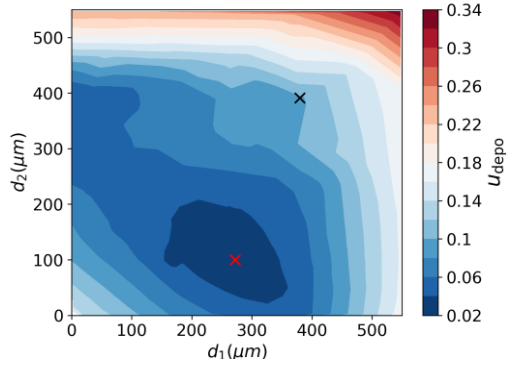

Figure 6: Laser irradiation uniformity corresponding to different laser aiming positions

## 4 Synthesized images

To provide more references for experiments, we employed post-processing methods to obtain synthesized images that represents the laser irradiation uniformity [23]. We adopted a synthesis strategy to estimate the instantaneous radiation flux $f(x,y)$. For a particular spatial section $X - O - Y$, we took the line integral value of the radiation power density $\rho(x,y,z)$ at each point in space as the true radiation flux. We divided the plane into a 512x512 grid and performed $10^8$ times random point counts based on the instantaneous radiation flux, ultimately presenting a normalized chromatic image of the radiation flux at each moment.

$$I(x, y) = \int \rho(x, y, z) dz$$

Figure 7 shows the synthetic X-ray radiation image for a typical case. Figure 7(a) presents a side-on view of the luminescent spherical cap target. Figure 7(b) provides face-on views of the luminescent spherical cap target at two different times. At the initial stage of laser irradiation, the spatial scale of the corona plasma is small, and the luminescent image of the spherical cap target can well reflect the overlay of the incident laser. In the later stage of laser irradiation, the fluid motion of the corona plasma, electron heat conduction, and spatial effects of laser energy deposition begin to have a significant impact. The luminescence of the corona plasma produces a notable smoothing effect on the incident laser, which no longer directly reflects the intensity overlay of the incident laser.

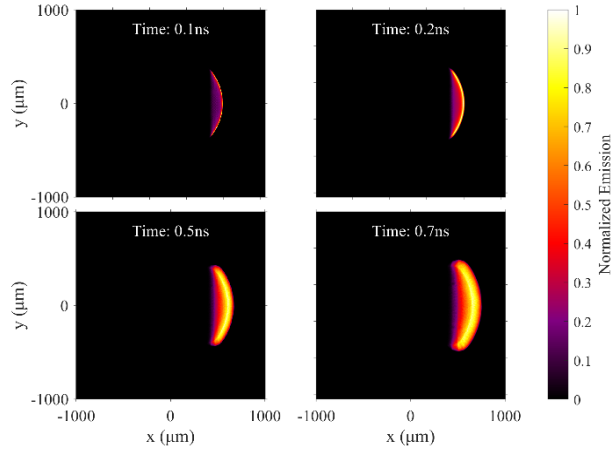

Figure 7: Side views of the X-ray split-magnitude phase at different time points

To study the relationship between plasma ablation and laser waveform, we also investigated the evolution of plasma ablation images driven by modulated laser waveforms. Figure 8 shows a typical laser waveform and the corresponding synthetic X-ray streak image. The laser energy is 8 kJ. The figure indicates a strong correlation between plasma ablation and the laser waveform. For example, when the laser starts irradiating, high-temperature plasma quickly appears on the surface of the spherical cap target. After the pre-pulse of laser, the plasma emission intensity weakens quickly. As the main laser arrives, the corona plasma expands more rapidly. When the laser ends, the plasma cools quickly. The X-ray radiation power curve in the figure shows the same trend.

Therefore, we believe that an X-ray streak camera can be used in experiments to infer the incident laser waveform.

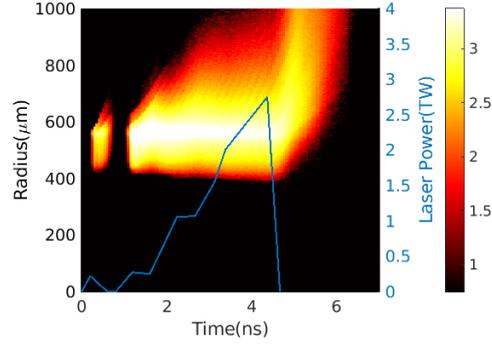

Figure 8: A modulated laser waveform and the synthetic image of an x-ray streak camera

## 5 Summary

We used the MULTI-3D radiation fluid program to study the interaction process between laser and target in the DCI scheme. Through a Bayesian optimization algorithm, we investigated the effects of the laser incident position of the laser beams on the uniformity of laser irradiation. It is found that the laser deposits energy throughout the corona region, but mainly near the critical surface. As the corona expands, the position of the critical surface slightly changes over time. If the laser is aimed at the top of the hemispherical target, the ablation pressure at the top will be greater than that at the edge, causing the spherical target to be flattened. Conversely, if the laser is aimed near the center of the target, the ablation pressure at the top will be too low to achieve spherical ablation drive.

It should be noted that since the ALE module for grid mapping in the MULTI-3D program is still under development. Therefore, this work only studied the laser ablation phase during the early implosion stage. In the future, we will use the MULTI-3D program with grid remapping capabilities to study the three-dimensional implosion process driven by lasers, providing more comprehensive theoretical support for laser fusion experiments.

## 6、Acknowledgment

This work was supported by the Strategic Priority Research Program of Chinese Academy of Sciences (Nos. XDA25051200, and XDA25010100), National Natural Science Foundation of China (125B100032)and AI for Science Program, Shanghai Municipal Commission of Economy and Informatization(2025-GZL-RGZN-BTBX-

02024).